\documentclass[prb,aps,twocolumn,amsmath,amssymb,floatfix,superscriptaddress]{revtex4}

\usepackage[dvips]{graphics}
\usepackage{color}

\definecolor{dred}{rgb}{0.75,0,0}

\usepackage[dvips]{graphics}
\usepackage{color}
\usepackage{soul}
\usepackage[colorlinks=true, citecolor=blue, urlcolor=blue ]{hyperref}

\begin{document}

\title{Antiferromagnetic diamond network as an efficient spin filter: Proposition of a spin-specific semi-conducting behavior}

\author{Debjani Das Gupta}

\affiliation{Physics and Applied Mathematics Unit, Indian Statistical Institute, 203 Barrackpore Trunk Road, Kolkata-700 108, India}

\author{Santanu K. Maiti}

\email{santanu.maiti@isical.ac.in}

\affiliation{Physics and Applied Mathematics Unit, Indian Statistical Institute, 203 Barrackpore Trunk Road, Kolkata-700 108, India}

\begin{abstract}

We propose, for the first time, that an array of diamond plaquettes, each possessing vanishing net magnetization, can achieve complete 
spin polarization over a broad bias window. Furthermore, this system can be utilized to realize spin-specific semiconducting behavior. 
We describe the antiferromagnetic diamond network within a tight-binding framework, where spin-dependent scattering arises due to the
interaction between itinerant electrons and local magnetic moments at different lattice sites. The mechanism underlying spin filtration 
relies on the specific arrangement of magnetic moments within individual plaquettes. We systematically investigate the spin polarization phenomenon under various input conditions, examining its dependence on network size, system temperature, and the magnetic flux threading 
each plaquette. Due to the network's geometry, we identify a sharply localized, highly degenerate energy level coexisting with conducting
states. By tuning physical parameters, a small energy gap can be established between these degenerate localized states and the conducting 
energy band, enabling spin-specific $p$-type and $n$-type semiconducting behavior. Our findings offer a novel approach for designing 
future spintronic devices based on similar antiferromagnetic networks.

\end{abstract}

\maketitle

\section{Introduction}

Antiferromagnetic (AFM) systems are emerging as promising candidates for next-generation spintronic
technology~\cite{xu,hu,jung,baltz,jung1,hoff,duine,nemec}. Over a decade ago, when researchers first incorporated electron spin alongside 
charge in microelectronic devices, ferromagnetic (FM) materials were the preferred choice~\cite{prinz,prinz1,wolf,zutic,fabian,jansen}. 
Ferromagnetic materials facilitate global spin polarization due to their long-range magnetic ordering. However, as technology advances, 
there is a growing demand for enhanced functionality, reduced power consumption, and miniaturized device architectures. In this context, 
AFM spintronic elements present several advantages over conventional FM materials. AFM systems can function as multilevel
switches~\cite{marrows, wadely} and, crucially, do not generate stray fringe fields, ensuring that their performance remains unaffected 
by external magnetic perturbations~\cite{jekins}. Moreover, they enable electronic writing pulse frequencies in the terahertz
regime~\cite{jung1,garello} and exhibit higher ordering temperatures compared to FM materials, allowing operation near room
temperature~\cite{jung,baltz}. Given these advantageous properties, AFM materials have been established as efficient components in 
spintronic applications~\cite{baltz,jung1,hoff}.

In spin-based electronic devices, a fundamental objective is to spatially separate spin-up and spin-down electrons along their transport 
path, enabling the utilization of spin degrees of freedom alongside charge transport. Achieving this requires the presence of spin-dependent
scattering within the system. When an electron moves at a sufficiently high velocity, the interaction between its angular momentum and 
spin effectively generates a magnetic field in its rest frame. This phenomenon, known as spin-orbit interaction (SOI), produces an effective
magnetic field that facilitates the separation of spin channels~\cite{soi,mdey}. In solid-state systems, two primary types of SOI are typically
considered: Rashba and Dresselhaus interactions. The Rashba interaction arises due to structural inversion asymmetry in the confining
potential~\cite{rashba}, while the Dresselhaus interaction originates from bulk inversion asymmetry~\cite{dres}. Although SOI-based
spin-dependent scattering mechanisms offer several advantages, a significant limitation is the generally weak spin-orbit coupling strength 
in most materials~\cite{strength}. This weak interaction reduces the efficiency of spin channel separation, making it challenging to 
achieve high spin currents and spin filtration over a reasonable bias window.

Given the inherent limitations of spin-orbit-coupled and ferromagnetic materials, recent breakthrough studies on antiferromagnetic (AFM) 
systems suggest that these materials offer a more suitable and reliable platform for the future development of spintronic devices with 
advanced functionalities. In general, the transmission spectra for spin-up and spin-down electrons are identical in AFM materials, 
resulting in no net spin current unless specific symmetry-breaking conditions are introduced. One approach to achieving spin polarization 
in AFM systems is to introduce asymmetric hopping in selective regions, thereby creating distinct transport environments for spin-up and
spin-down electrons. Alternatively, a disparity in onsite potential can enable electrons of different spins to propagate through separate 
energy channels~\cite{dasg}. Various methods have been proposed to induce asymmetric hopping in different geometric configurations.
Additionally, chirality combined with an external electric field can create spin-dependent potential energy variations in AFM molecular
systems~\cite{dasg1,dasg2}. The efficiency of spin filtration is enhanced in AFM helical structures with long-range hopping order. 
However, in the absence of helicity or an applied electric field, spin separation is completely suppressed. 

\begin{figure}[ht]
\centering
\resizebox{8.5cm}{1.75cm}{\includegraphics{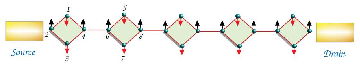}}
\caption{(Color online). Schematic of the junction setup, where an antiferromagnetic diamond network is coupled to source and drain electrodes.}
\label{diamond}
\end{figure}

We seek a system where spin channel separation can be achieved without the application of an external electric field, even in the presence 
of symmetric hopping. In this work, we demonstrate that a mesoscopic chain with a specific geometric structure—comprising multiple loops 
and exhibiting zero net magnetization—can generate fully spin-polarized current (see Fig.~\ref{diamond}). The proposed antiferromagnetic 
network consists of multiple diamond plaquettes, each containing four magnetic sites arranged such that the net magnetization of each 
plaquette remains zero. The motivation for considering this geometry is multifaceted. 
First, looped structures inherently exhibit nontrivial
quantum interference effects due to the presence of multiple electronic transport paths~\cite{bercioux,aharony,sil,pal,saha}, distinguishing
them from conventional loopless systems. Second, by ensuring that each diamond plaquette has zero net magnetization, the symmetry between
spin-up and spin-down Hamiltonians is naturally broken, enabling spin-selective electron transmission through the network. Third, this 
specific geometry gives rise to a sharply localized energy level coexisting with conducting energy states. Notably, this localized level 
is highly degenerate, facilitating spin-specific semiconducting behavior. These findings suggest a new design paradigm for efficient 
electronic and spintronic devices, where spin channel separation can be realized without explicit symmetry breaking or external field
application. Instead, multiple AFM loop substructures inherently enable spin-selective transport, paving the way for novel spintronic
architectures.

We simulate the quantum system within a tight-binding (TB) framework. The presence of a sharply localized energy level 
is analytically determined from the energy dispersion relation of the network. To investigate spin-selective electron transport, we couple 
a finite-sized diamond network to two contact electrodes and compute the transmission probabilities using the Green's function
formalism. The spin-dependent currents are then evaluated using the Landauer-Büttiker formula, allowing us to determine the spin polarization
coefficient. To explore the tunability of spin filtration efficiency, we introduce a finite magnetic flux through each plaquette and observe
that the phase of spin polarization can be selectively modified. Notably, the sharply localized energy level vanishes when a magnetic flux
$\phi$ is applied but re-emerges at half the flux quantum ($\phi_0/2$). The spin-selective semiconducting behavior is directly linked to this
degenerate localized energy level. By appropriately adjusting the Fermi energy, the system can exhibit either p-type or n-type spin-dependent
semiconducting properties. Furthermore, we critically analyze the effects of system temperature and size to ensure a comprehensive 
understanding of the proposed model.

The remainder of this paper is organized as follows. Section II describes the junction setup, TB Hamiltonian, and theoretical
framework for the calculations. Section III presents the results and is divided into three parts. First, we analytically derive 
the energy band structure and examine the behavior of localized states in the presence of a magnetic flux. Next, we discuss possible 
tuning mechanisms for spin polarization. Finally, the last part explores spin-dependent semiconducting behavior. A summary of our 
findings is provided in Sec. IV.

\section{Model and Theoretical Formalism}

\subsection{Model Description and Tight-Binding Hamiltonian}

Let us begin with the schematic representation of the spin-polarized setup shown in Fig.~\ref{diamond}. The central system consists 
of an array of magnetic diamonds, with neighboring diamonds connected diagonally by a single bond. This array is attached to two 
non-magnetic electrodes at its ends: a source (S), where electrons are injected, and a drain (D), where electrons are collected. 
The electrodes are considered to be one-dimensional, perfect, and non-magnetic. The different components of the junction setup, viz, 
the AFM diamond network, electrodes and their coupling with the network, are illustrated as follows.

Within the diamond-like plaquette structure, two distinct atomic positions exist. At sites $1$ and $3$ (see Fig.~\ref{diamond}), 
atoms have two
nearest neighbors, these are labeled as A-type atoms. In contrast, sites $2$ and $4$ host atoms with three nearest neighbors, referred to 
as B-type atoms. Each site A contains an effective magnetic moment along the $-Z$ direction, while site B has an effective magnetic moment in the opposite direction i.e., along $+Z$. The condition of vanishing net magnetization can be achieved in different ways, however, the specific arrangement of local moments is crucial for ensuring spin-selective electron transmission, as will be evident from our forthcoming discussion. It is important to note that our chosen AFM network is not a `classical' one, thus the local 
moments are not classical spins. Instead, each site of the network can be described as an effective magnetic moment aligned along $+Z$ or 
$-Z$. These magnetic moments are treated as static Zeeman-like fields. We assume that the magnitudes of these fields are identical,
denoted by $h$, but their signs are opposite depending on their alignments along $\pm Z$.  The itinerant electrons interact with these local 
magnetic moments. For any $i$-th site where the effective magnetic moment is aligned along $+Z$, an up-spin electron experiences an 
effective potential $\epsilon_i-h$, while a down-spin electron experiences $\epsilon_i+h$~\cite{strength,patra,sarkar}. Here, $\epsilon_i$ represents the site energy in the absence of magnetic scattering. These potentials are interchanged for the $-Z$ configuration.

Our system is not a classical spin blockage problem, rather it is a quantum interference problem. It is also free
from dynamic coupling, Hund's rule exchange interaction and spin-dependent hopping amplitude.

Considering the spin-dependent scattering mechanism described above along with the intra- and inter-diamond hopping 
integrals, the general form of TB Hamiltonian of the AFM diamond network can be written as:
\begin{equation}
\boldsymbol{\mathcal{H}_{\mbox{\tiny AFM}}} = \sum_i \left[ \boldsymbol{c}_i^\dagger \left(\boldsymbol{\epsilon}_i - \vec{\boldsymbol{h}}_i \cdot \vec{\boldsymbol{\sigma}} \right) \boldsymbol{c}_i + \boldsymbol{c}_{i+1}^\dagger \boldsymbol{t}_i \boldsymbol{c}_i + h.c. \right],
\label{tb}
\end{equation}
where $\boldsymbol{c}_i^\dagger = \begin{pmatrix} c_{i\uparrow}^\dagger & c_{i\downarrow}^\dagger \end{pmatrix}$, with $c_{i\sigma}^\dagger$ 
and $c_{i\sigma}$ being the creation and annihilation operators at site $i$ for spin $\sigma = (\uparrow, \downarrow)$, respectively. 
Any arbitrary orientation of $\vec{\boldsymbol{h}}_i$ is characterized by a polar angle $\theta_i$ and an azimuthal angle $\varphi_i$ in spherical coordinates. Thus, the modified site energy matrix can be written as:
\begin{equation}
\boldsymbol{\epsilon}_i - \vec{\boldsymbol{h}}_i \cdot \vec{\boldsymbol{\sigma}} =
\begin{pmatrix}
\epsilon_i - h\cos\theta_i & -h\sin\theta_i e^{-j\varphi_i} \\
-h\sin\theta_i e^{j\varphi_i} & \epsilon_i + h\cos\theta_i
\end{pmatrix}.
\label{site}
\end{equation}
Here, the non-magnetic potential energy $\epsilon_i$ is assumed to be identical for both spin-up and spin-down electrons at any given site, 
and as stated earlier, the magnitudes of $h_i$'s are identical at all sites.  

The hopping energy matrix in Eq.~\ref{tb} is given by:
\begin{equation}
\boldsymbol{t}_i = \text{diag}(t_i, t_i),
\end{equation}
where $t_i = t$ within a diamond plaquette and $t_i = t^{\prime}$ between adjacent diamond plaquettes. Four our chosen quantum AFM system,
electrons hop with a constant hopping strength in each plaquette, independent of the orientation of $\vec{h}_i$. The effect of spin-dependent 
scattering enters into the Hamiltonian via on-site potentials, not via the modification of the hopping.

The TB Hamiltonian for the side-attached non-magnetic electrodes (taken in the form of a 1D chain) is written as: 
\begin{equation}
\boldsymbol{\mathcal{H}_{\mbox{\tiny S(D)}}} = \sum_i \left[ \boldsymbol{c}_i^\dagger \boldsymbol{\epsilon}_0 \boldsymbol{c}_i + \boldsymbol{c}_{i+1}^\dagger \boldsymbol{t}_0 \boldsymbol{c}_i + h.c. \right],
\label{elec}
\end{equation}
where $\boldsymbol{\epsilon}_0 = \text{diag}(\epsilon_0, \epsilon_0)$ and $\boldsymbol{t}_0 = \text{diag}(t_0, t_0)$. The TB parameters, 
$\epsilon_0$ and $t_0$, represent the site energy and nearest-neighbor hopping strength, respectively, for the electrodes.

To form the nanojunction, we couple the two end sites of the diamond network to source and drain. We define these coupling 
strengths by the parameters $\tau_S$ and $\tau_D$, respectively. The electrodes can also be connected to two other sites, but in the present discussion we select this specific junction configuration such that the arm lengths with respect to source and drain are equal.

\subsection{Theoretical Formalism: ADOS, Transmission probability, Current, and Spin polarization}

For our study, we need to calculate different physical quantities those are interrelated to each other, and we compute them using the 
well-known Green's function formalism~\cite{datta}.

The average density of states (ADOS) is obtained by taking the imaginary part of the trace of the retarded Green's function:
\begin{equation}
\rho=-\left(\frac{1}{\pi N_s}\right)\textrm{Im}[\textrm{Tr}[\boldsymbol{\mathcal{G}^r}]],
\label{dos1}
\end{equation}
where $N_s$ is the total number of sites in the diamond network. The total number of diamond plaquettes is denoted by the parameter 
$N$ ($N_s=4N$). The factor $\boldsymbol{\mathcal{G}^r}$, referred to as retarded Green's function, is defined as:
\begin{equation}
\boldsymbol{\mathcal{G}^r} = \left(E\boldsymbol{I} - \boldsymbol{\mathcal{H}_{\mbox{\tiny AFM}}} - \boldsymbol{\Sigma_S} - \boldsymbol{\Sigma_D}\right)^{-1},
\label{grf}
\end{equation}
where $\boldsymbol{I}$ is the identity matrix, and, $\boldsymbol{\Sigma_S}$ and $\boldsymbol{\Sigma_D}$ are the self-energies of the 
source and drain leads, respectively. The self-energies contain all the information of the electrodes and their coupling with the diamond 
network. The non-zero element in $\boldsymbol{\Sigma_S(D)}$ depends on the coupling point of S(D) with the network, and is
expressed as: $\tau_{S(D)}^2/2t_0^2\left[(E-\epsilon_0)-j \sqrt{4t_0^2-(E-\epsilon_0)^2}\right]$, where $j=\sqrt{-1}$. A detailed 
derivation of it is found in Ref.~\cite{datta}. When the density of states is calculated solely for the conductor itself, i.e., in the absence of the electrodes, the self-energy terms are omitted in the expression of Green's function. 

The spin-dependent transmission probability is obtained from the expression:
\begin{equation}
T_{\sigma \sigma^{\prime}} = \textrm{Tr}[\boldsymbol{\Gamma_{S\sigma} \mathcal{G}^r \Gamma_{D\sigma{\prime}} \mathcal{G}^a}],
\end{equation}
where $\boldsymbol{\mathcal{G}^a}$ is the advanced Green's functions and it is connected to the 
retarded Green's function as $\boldsymbol{\mathcal{G}^a}=(\boldsymbol{\mathcal{G}^r})^\dagger$.
$\boldsymbol{\Gamma_{S\sigma}}$ and $\boldsymbol{\Gamma_{D\sigma{\prime}}}$ are the coupling matrices, and they are related to the
self-energies as $\boldsymbol{\Gamma_{S(D)}} = -2 \textrm{Im}[\boldsymbol{\Sigma_{S(D)}}]$. After evaluating the individual components, 
we compute the net up spin transmission probability using the relation $T_{\uparrow}=T_{\uparrow\uparrow}+T_{\downarrow\uparrow}$. 
Similarly, by adding the other two components, we determine the net down spin transmission probability as  $T_{\downarrow}=T_{\downarrow\downarrow}+T_{\uparrow\downarrow}$.

We calculate the spin-resolved current using the Landauer-B\"uttiker formalism~\cite{sarkar,datta}. The total transmission probability
$T_\sigma$ is integrated over the full allowed energy window to obtain the current at a finite temperature $\mathcal{T}$. Under an 
applied voltage bias $V$ between the two electrodes, the spin-dependent current is given by:
\begin{equation}
I_\sigma = \left(\frac{e}{h}\right) \int T_\sigma \left(f_S - f_D\right) dE.
\end{equation}
Here, $f_{S(D)}$ represents the Fermi function of the source (drain), expressed as:
\begin{equation}
f_{S(D)} = \left[1 + \exp\left(\frac{E - \mu_{S(D)}}{k_B \mathcal{T}}\right)\right]^{-1},
\end{equation}
where the electrochemical potentials of the electrodes are related to the Fermi energy $E_F$ of the system as $\mu_{S(D)} = E_F \pm eV/2$.

The spin polarization is defined as the ratio of spin current to charge current:
\begin{equation}
P = \frac{I_\uparrow - I_\downarrow}{I_\uparrow + I_\downarrow}.
\end{equation}
A value of $P = \pm 1$ indicates complete spin polarization, while $P = 0$ corresponds to equal spin-up and spin-down currents, 
resulting in zero net spin current and polarization.

\section{Results and Discussion}

In what follows, we present our results, organized into different sub-sections. Each result is critically analyzed to facilitate a clear understanding. Throughout the description, all the energies are measured in units of eV.

\subsection{Spectral Analysis and Localization Behavior}

We first examine the band structure of an isolated diamond network i.e., when it is not coupled to source and drain electrodes. 

\vskip 0.2cm
\noindent
$\blacksquare$ {\bf Non-magnetic diamond network}:  
We begin with the spinless case, where all sites are nonmagnetic. In the subsequent part, we extend these results 
\begin{figure}[ht]
{\centering\resizebox*{8cm}{3cm}{\includegraphics{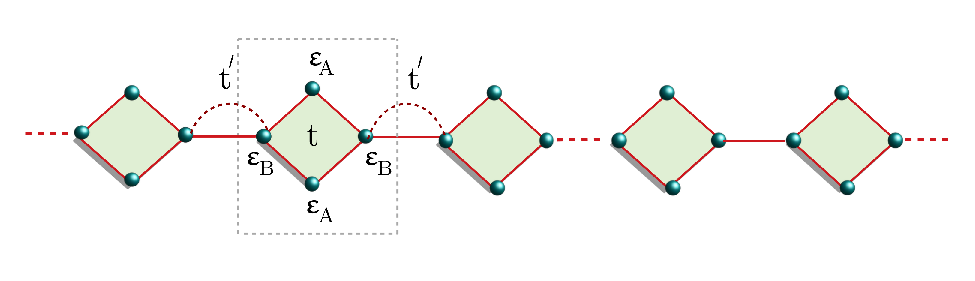}}\par}
\caption{(Color online). Infinite array of diamond plaquettes. We consider the spinless case, where two types of hopping energies are 
present: $t$ within a plaquette and $t^{\prime}$ between adjacent plaquettes. The unit cell is indicated in the figure.}
\label{model}
\end{figure}
in presence of spin-dependent scattering, incorporating our actual antiferromagnetic model.

From the energy dispersion relation, we can identify the localization and conduction bands, which later help in determining 
\begin{figure}[ht]
	{\centering\resizebox*{7.5cm}{5cm}{\includegraphics{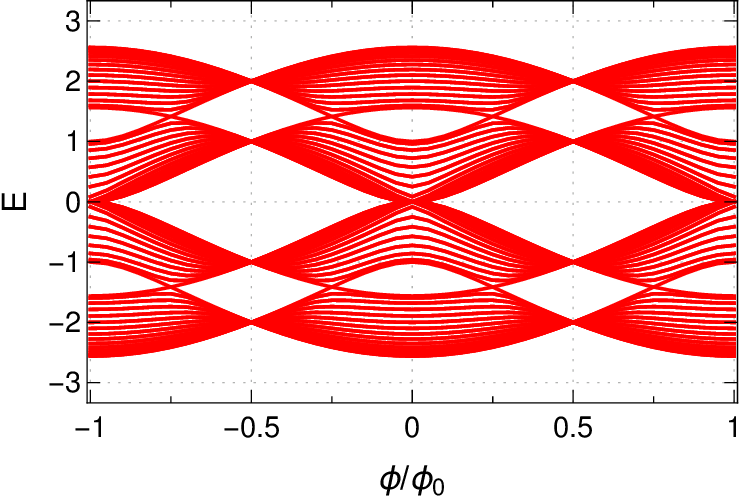}}\par}
	\caption{(Color online). Variation of energy eigenvalues with magnetic flux. Four energy bands are observed, with band gaps appearing at specific values of $\phi$. We consider $N=20$ plaquettes under periodic boundary conditions to approximate an infinite diamond chain. The parameters used are $\epsilon_A = \epsilon_B=0$ and $t=t^{\prime}=1$.}
	\label{ephi}
\end{figure}
the energy window for either spin-up or spin-down electrons. Each unit cell consists of four atomic sites, as shown in Fig.~\ref{model}. There are two types of sites, labeled A and B, with intracell hopping energy $t$ and intercell hopping energy $t^{\prime}$. The effective site energy matrix and hopping matrix for a unit cell are given by
\begin{eqnarray}
\boldsymbol{\epsilon_{\text{eff}}} =
\begin{pmatrix}
\epsilon_A & t & 0 & t\\
t & \epsilon_B & t & 0\\
0 & t & \epsilon_A & t \\
t & 0 & t & \epsilon_A 
\end{pmatrix}, \quad
\boldsymbol{t_{\text{eff}}} =
\begin{pmatrix}
0 & 0 & 0 & 0\\
0 & 0 & 0 & 0\\
0 & 0 & 0 & 0\\
0 & t^{\prime} & 0 & 0
\end{pmatrix}.
\end{eqnarray}
Furthermore, we can generalize this system to the presence of a magnetic flux. If a uniform magnetic field $\vec{B}$ is 
\begin{figure}[ht]
{\centering\resizebox*{7cm}{11cm}{\includegraphics{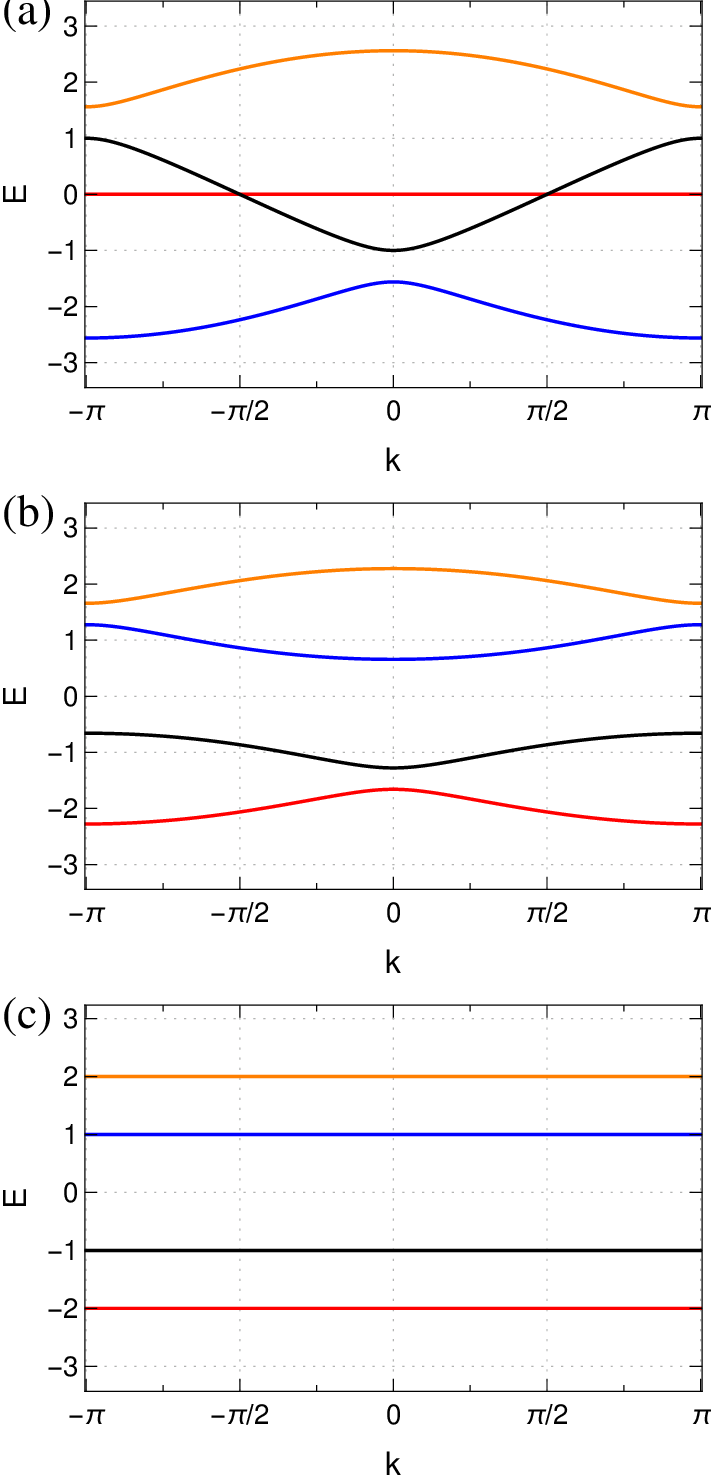}}\par}
\caption{(Color online). $E$-$k$ relation for different magnetic flux values: (a) $\phi=0$, (b) $\phi=0.35\phi_0$, and (c) $\phi= 0.5\phi_0$. Different colors represent different energy bands. The parameters used are identical to those in Fig.~\ref{ephi}.}
\label{ek}
\end{figure}
applied along the $Z$-axis, threading a flux $\phi$ through each loop, it induces a phase difference as electrons hop between sites within a diamond plaquette~\cite{saha,gefen}. This modifies the hopping energy $t$, leading to a transformation:
$$ t \rightarrow t e^{j\eta}, \quad \text{where } \eta = \frac{\pi \phi}{2\phi_0}.$$
Here, $\phi$ is measured in units of the flux quantum $\phi_0 = hc/e$.

To determine the energy dispersion relation, we solve the determinant equation:
\begin{equation}
\left| E \boldsymbol{I} - \boldsymbol{\epsilon_{\text{eff}}} - e^{jk} \boldsymbol{t_{\text{eff}}} - e^{-jk} \boldsymbol{t_{\text{eff}}}^{\dagger} \right| = 0.
\end{equation}

Several interesting features arise in the energy bands as the flux $\phi$ varies. These effects are discussed in detail below.

By solving the above equation, we obtain the $E$-$k$ relation for the spinless case in the presence of magnetic flux:
\begin{eqnarray}
\left(E - \epsilon_A \right)^2 \left[\left(E -\epsilon_B \right)^2 - t^{{\prime}^2} \right] - 4t^2\left(E - \epsilon_A\right) 
\left[\left(E - \epsilon_B\right) \right. \nonumber \\
\left.\quad  + t^{\prime}\cos 2\eta \cos k\right] + 2t^4 (1 - \cos 4\eta) = 0.\quad
\label{ek1}
\end{eqnarray}

These energy eigenvalues (analytical expressions for various $E$-$k$ relations are provided in the Appendix~\ref{app} for clarity) 
exhibit a periodicity of $\phi_0$, as illustrated in Fig.~\ref{ephi}. Four energy bands are observed, with some bands touching at
\begin{figure}[ht]
{\centering\resizebox*{7cm}{11cm}{\includegraphics{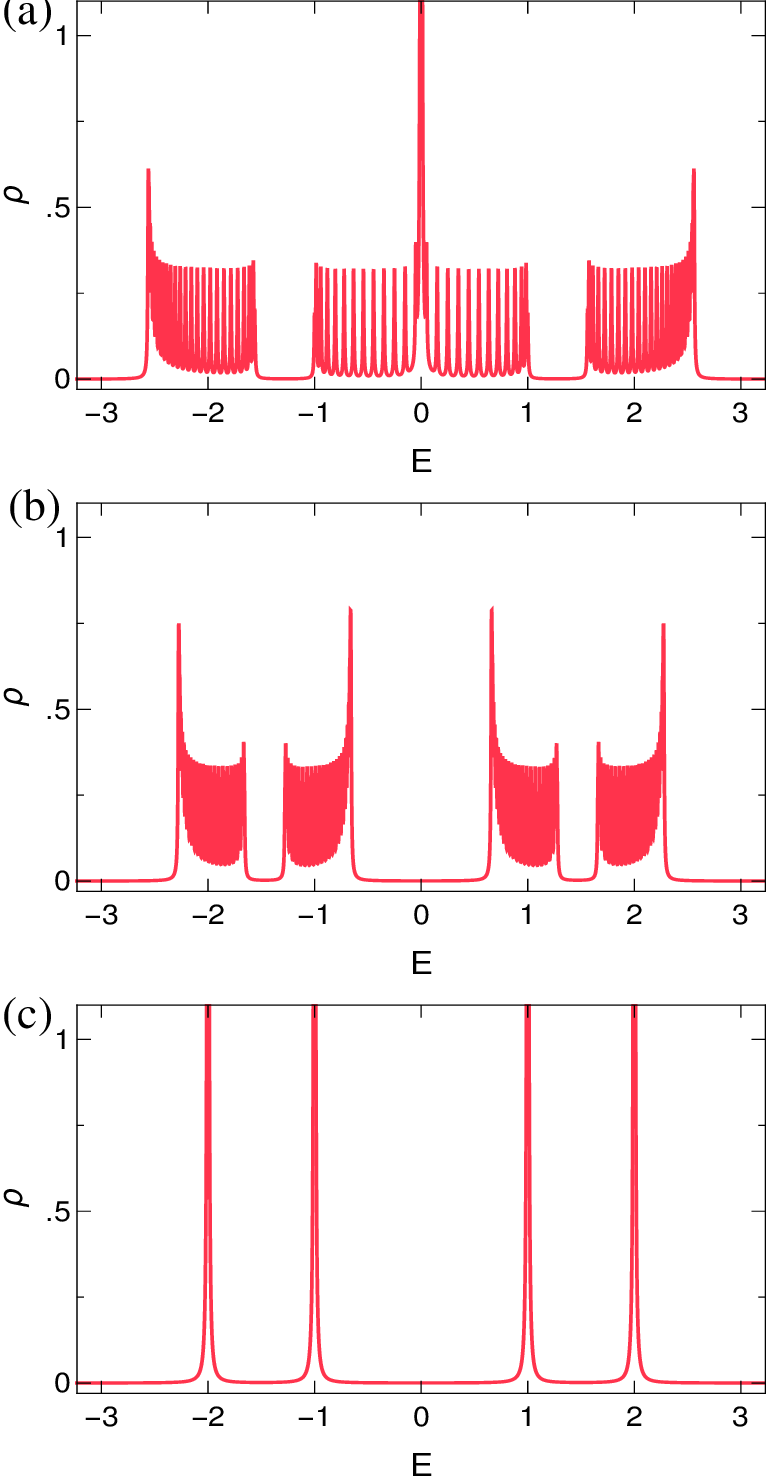}}\par}	
\caption{(Color online). Average density of states for the same flux values as in Fig.~\ref{ek}. We consider $N=50$ plaquettes, with parameters identical to those in Fig.~\ref{ephi}.}
\label{dos}
\end{figure}
$\phi/\phi_0 = 0$ or $1$. The band gap maximizes at $\phi/\phi_0 = 0.5$. To analyze localization behavior, we consider three distinct 
cases:

\begin{itemize}
\item \textbf{Case 1:} $\phi = 0$ (no magnetic flux)
	
In the absence of a magnetic flux, the phase factor vanishes, simplifying Eq.~\ref{ek1}. In this case, there always exists a solution 
$E = \epsilon_A$, independent of $k$, corresponding to a localized state (see Eq.~\ref{ap3}). This is depicted in Fig.~\ref{ek}(a), where a 
straight (red) line at $E=0$ indicates localization. The same is reflected from the profile of density of states as shown in Fig.~\ref{dos}(a), where a sharp peak appears.
	
\item \textbf{Case 2:} $0<\phi<\phi_0/2$ (intermediate magnetic flux)
	
At intermediate values of $\phi$, all energy eigenvalues depend on $k$, eliminating localized states (see Eqs.~\ref{ap1} and \ref{ap2}).
Figure~\ref{ek}(b) shows the dispersion relation for $\phi = 0.35\phi_0$, where no localized states appear. The corresponding density 
of states in Fig.~\ref{dos}(b) further confirms their absence.
	
\item \textbf{Case 3:} $\phi = \phi_0/2$ (half flux-quantum)
	
At $\phi = \phi_0/2$, localized bands reappear. Unlike the zero-flux case, four distinct localized levels emerge at $E= -2, -1, 1$, and $2$, 
as observed in Fig.~\ref{ek}(c), following Eqs.~\ref{ap7} and \ref{ap8}. The degeneracy of these states matches the number of diamond
plaquettes, $N$, under periodic boundary conditions. The same is reflected in Fig.~\ref{dos}(c).
\end{itemize}

\vskip 0.2cm
\noindent
$\blacksquare$ {\bf AFM diamond network}:
Now we analyze the spin-dependent energy band structure of our chosen antiferromagnetic system.  
\begin{figure}[ht]
\centering
\resizebox*{8cm}{7.5cm}{\includegraphics{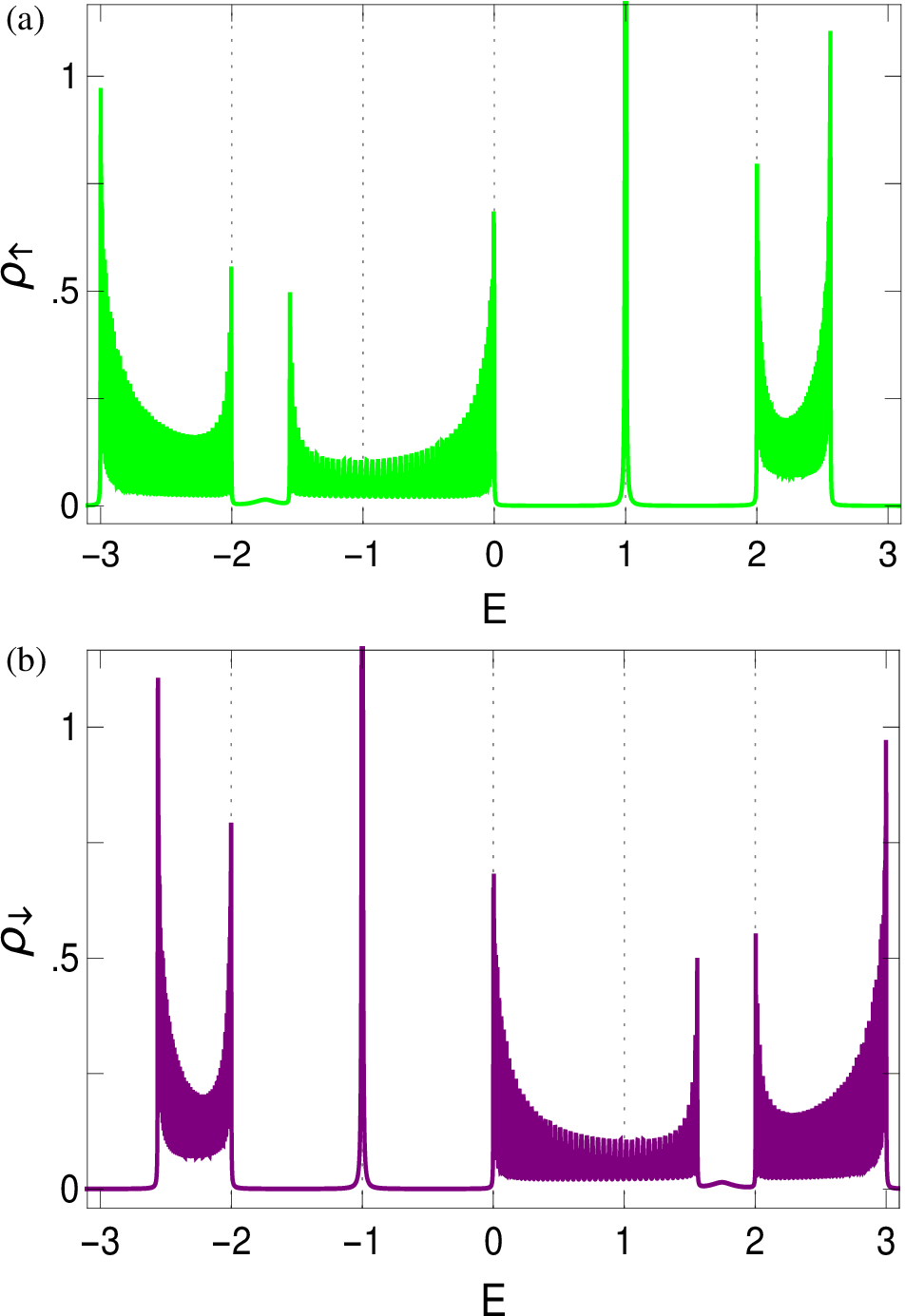}}
\caption{(Color online). ADOS for (a) up spin and (b) for down spin in an AF diamond network with $N = 50$. The sharp peak at $E = 1$ in (a) corresponds to a localized state for up spin, while in (b), the localized state for down spin is at $E = -1$. A finite energy gap between the conducting band and the localized state can lead to a spin-selective transport.}
\label{density}
\end{figure}
Here, we consider the case with zero magnetic flux. The influence of magnetic flux will be addressed in the following sub-sections. 
Understanding the spin-dependent energy 
bands is crucial for realizing efficient spintronic devices. In our model, the magnetic moments are oriented along either the $+Z$ or $-Z$ 
axis, i.e., $\theta_i = 0$ or $\pi$, with $\varphi_i = 0$. This results in diagonal onsite and hopping energy matrices, allowing us to 
decompose the Hamiltonian into two independent spin sub-spaces. In the AF network (Fig.~\ref{diamond}), sites A (1,3,5,7) have magnetic 
moments $\vec{h} (h, \pi, 0)$ along $-Z$, while sites B (2,4,6,8) have $\vec{h} (h, 0, 0)$ along $+Z$. The spin-dependent onsite potentials are:
\begin{align}
\epsilon_{A\uparrow} &= \epsilon + h, & \epsilon_{B\uparrow} &= \epsilon - h, \\
\epsilon_{A\downarrow} &= \epsilon - h, & \epsilon_{B\downarrow} &= \epsilon + h.
\end{align}
The dispersion relations are:
\begin{eqnarray}
\left(E - \epsilon - h \right) \left[ \left(E - \epsilon + h \right) \left\{ \left(E - \epsilon \right)^2 - h^2 - 4t^2 \right\} 
\right. \nonumber \\
\left. - 4t^2 t^{\prime} \cos k \right] = 0,
\label{up}
\end{eqnarray}
for up spin electrons, and
\begin{eqnarray}
\left(E - \epsilon + h \right) \left[ \left(E - \epsilon - h \right) \left\{ \left(E - \epsilon \right)^2 - h^2 - 4t^2 \right\} 
\right. \nonumber \\
\left. - 4t^2 t^{\prime} \cos k \right] = 0,
\label{dn}
\end{eqnarray}
for down spin electrons.
Equation~\ref{up} shows that up spin electrons form a localized level at $E_\uparrow = \epsilon + h$, while Eq.~\ref{dn} indicates that down spin electrons localize at $E_\downarrow = \epsilon - h$. If a finite gap exists between the conducting and localized states, the Fermi energy can be tuned to enable transport of only one spin species, achieving spin filtering.
 
To have a clear understanding of different sub-bands, we compute ADOS for a finite AF network. For each spin case, the ADOS is 
determined by the relation
\begin{equation}
\rho_\sigma = -\frac{1}{\pi N_s} \textrm{Im} [\textrm{Tr} [\boldsymbol{\mathcal{G}^r_\sigma}]],
\end{equation}
where the retarded Green’s function is defined as 
\begin{equation}
\boldsymbol{\mathcal{G}^r_\sigma} = \left( \boldsymbol{E} - \boldsymbol{\mathcal{H}_\sigma} - \boldsymbol{\Sigma_{S\sigma}} - 
\boldsymbol{\Sigma_{D\sigma}} \right)^{-1}.
\end{equation}
In Fig.~\ref{density}, we present the ADOS plots for up and down spin cases, using the parameters $\epsilon = 0$, $t = t^{\prime} = 1$, 
and $h = 1$. From the spectra, we find that localized states for up and down spin electrons appear at two distinct energies. A significant mismatch between the energy channels of up and down spins is also evident. By appropriately choosing the Fermi energy, one spin species can be completely blocked while the other propagates freely, leading to full spin polarization. This behavior is observed across multiple energy regions.

\vskip 0.2cm
\noindent
{\em Role of arrangement of local magnetic moments}: Here it is crucial to point out that, in order to achieve
\begin{figure}[ht]
\centering
\resizebox*{5cm}{5cm}{\includegraphics{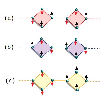}}
\caption{(Color online). Some possible alternative configurations of local magnetic moments in diamond plaquettes, other than what is shown in Fig.~\ref{diamond}. In each of these configurations, the net magnetization is zero.}
\label{ang}
\end{figure}
spin channel separation and spin-selective electron transmission, the specific arrangement of local magnetic moments at different lattice sites plays a vital role. More precisely, we require a configuration in which the symmetry between the up and down spin sub-Hamiltonians is broken. From Fig.~\ref{diamond}, we observe that an up moment is surrounded by one up and two down moments, whereas a down moment is surrounded by two up moments. This asymmetry leads to a clear imbalance that breaks the symmetry between the two spin-dependent energy channels. However, if the local moments are arranged such that the neighboring moments around up and down moments are equivalent, the symmetry between the up and down spin sub-Hamiltonians is restored, and spin channel separation no longer occurs. To illustrate this point, in Fig.~\ref{ang}, we show several such symmetric arrangements. In all these configurations, the neighboring moments around up and down moments are equivalent. As a result, spin-selective transport phenomena are not expected in these cases.

\subsection{Spin-Selective Transmission, Filtration, and Related Issues}

Following the above analysis of spectral behavior, in this sub-section we focus on spin-selective transmission probabilities, spin currents, 
the degree of spin filtration, and related aspects. For these calculations, some physical parameters are kept fixed, as specified below.
In the side-attached electrodes, the site energy is set to $\epsilon_0=0$ and the hopping strength is $t_0=2$. In the AFM network, the site energy is $\epsilon_i=0$, and unless specified the spin-dependent scattering factor is taken as $h=1$ and the azimuthal and polar angles are set to zero. The coupling strengths between the electrodes and the diamond network are $\tau_S=\tau_D=1$.

\begin{figure*}[ht]
{\centering\resizebox*{16cm}{4cm}{\includegraphics{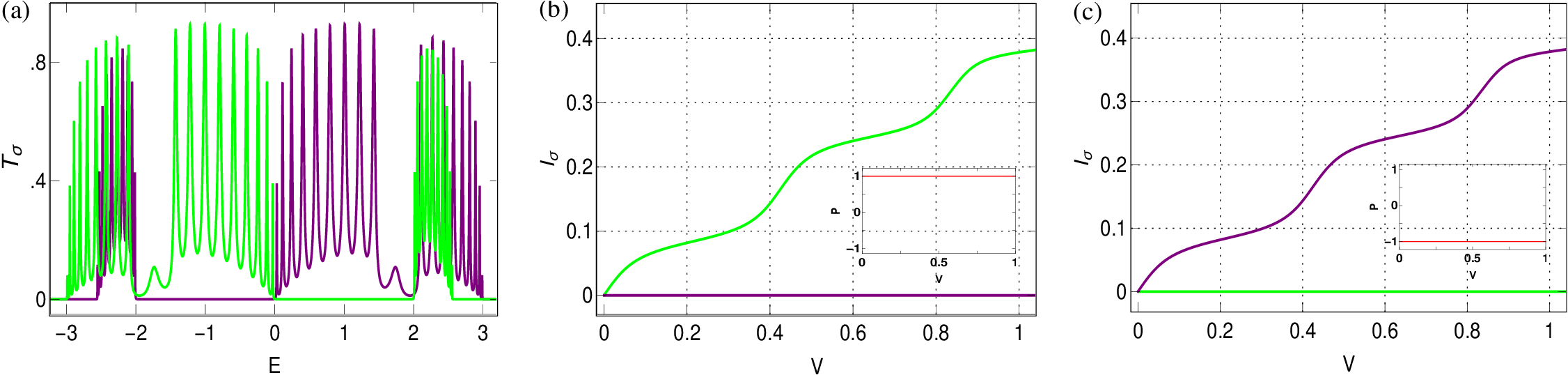}}\par}
\caption{(Color online). (a) Spin-dependent transmission probabilities as function of energy for the network with number of plaquettes $N=10$. 
A finite mismatch between the two spin bands is clearly observed, resulting in different energy zones where one spin is completely 
blocked while the other is transmitted. The spin-dependent currents at $\mathcal{T}=10K$ for two different Fermi energies, $E_F=-1$ and 
$E_F=1$, are shown in (b) and (c), and, the insets shows the polarization curve with bias voltage at these Fermi energies respectively. 
The green curve represents the up spin, while the purple curve represents the down spin in (a)–(c).}
\label{current}
\end{figure*}

Let us begin with Fig.~\ref{current}, which presents the spin-dependent transmission probabilities, currents and spin polarization 
for an AFM diamond network with $N=10$.
Figure~\ref{current}(a) shows the variation in up and down spin transmission probabilities, clearly revealing their mismatch.
Over a wide energy region, electrons of one spin propagate, while those of the other spin are completely blocked. In the transmitting 
zone, multiple resonant peaks are observed, all of which are directly associated with the available energy channels of the conductor. 
The widths of these peaks are determined by the coupling strength between the conductor and the side-attached electrodes. Under weak coupling, 
the peaks are sharp, whereas they broaden with increasing coupling strength--an effect well known in the literature.
Figures~\ref{current}(b) and (c) present the spin-dependent currents and the corresponding spin polarizations for two distinct Fermi 
energies: $E_F=-1$ and $E_F=1$, respectively. When $E_F=-1$, only the up-spin current is observed within the chosen bias window, while 
the down-spin current is completely suppressed, reflecting the nature of the transmission profiles. As a result, $100\%$ spin polarization 
is achieved, as shown in the inset of Fig.~\ref{current}(b). The step-like behavior of the current arises from the sharpness of the 
resonant transmission peaks. An exactly opposite scenario occurs when the Fermi energy is shifted to $E_F=1$: only the down-spin current 
is present, leading to $-100\%$ spin polarization (Fig.~\ref{current}(c) and its inset), in agreement with the transmission curves. 
Thus, by selectively tuning the Fermi energy, complete spin polarization can be achieved and sustained over a large bias window.

Next, we examine the angular dependence of spin transmission. The polar angle for site A is denoted as $\theta$, while for site B, 
it is $\pi + \theta$. This configuration ensures an antiferromagnetic arrangement in each plaquette.
\begin{figure}[ht]
{\centering\resizebox*{6.5cm}{6.5cm}{\includegraphics{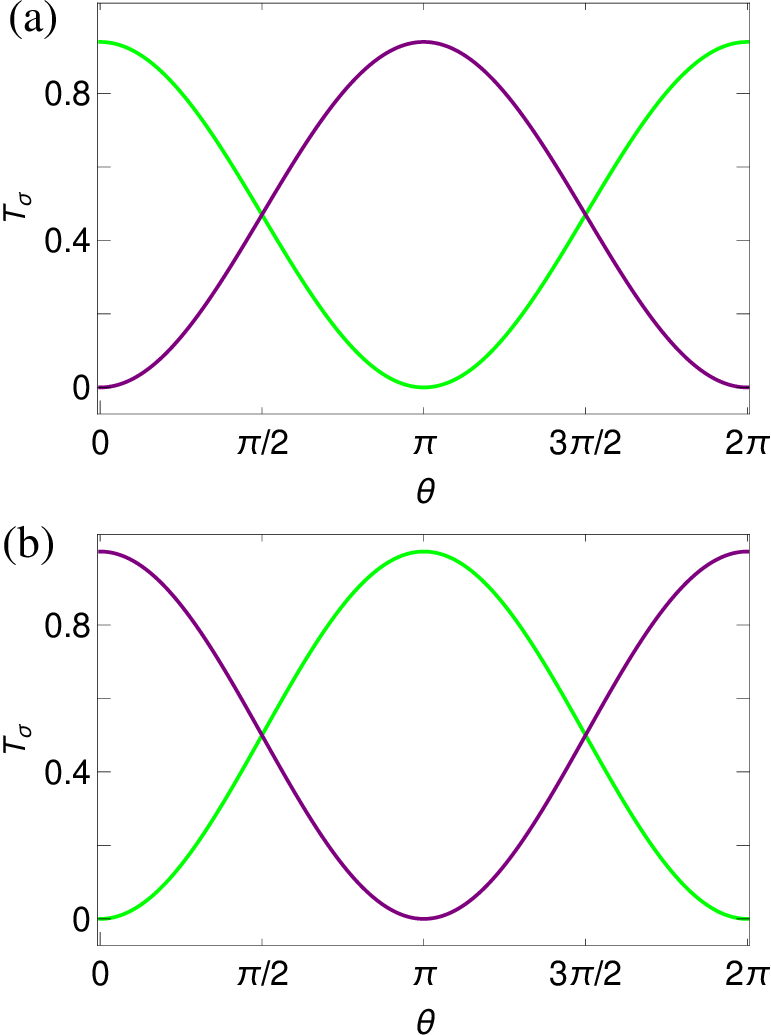}}\par}
\caption{(Color online). Transmission probabilities, at a particular energy $E$, as a function of the polar angle $\theta$. The magnetic moments at A-type sites are oriented at an angle $\theta$, while at B-type sites, they are $\pi + \theta$. The energy is set at $E = -1$ in (a) and at $E = 1$ in (b). The color convention follows that of Fig.~\ref{current}, where the green curve represents up-spin and the purple curve represents down-spin. Here, $N=10$.}
\label{theta}
\end{figure}
We analyze the effect of magnetic moment rotation on spin filtration by calculating the transmission probability, spin-dependent current, 
and polarization. Figures~\ref{theta}(a) and (b) show the transmission probabilities $T_\uparrow$ (green) and $T_\downarrow$ (purple) for 
two different energies, $E = -1$ and $E = 1$. These curves exhibit sinusoidal variations, with the maximum phase difference 
occurring when the polar angle is an integer multiple of $\pi$. The same pattern is observed in the spin-dependent currents shown in
Figs.~\ref{rotate}(a) and (c).
\begin{figure}[ht]
{\centering\resizebox*{8cm}{6cm}{\includegraphics{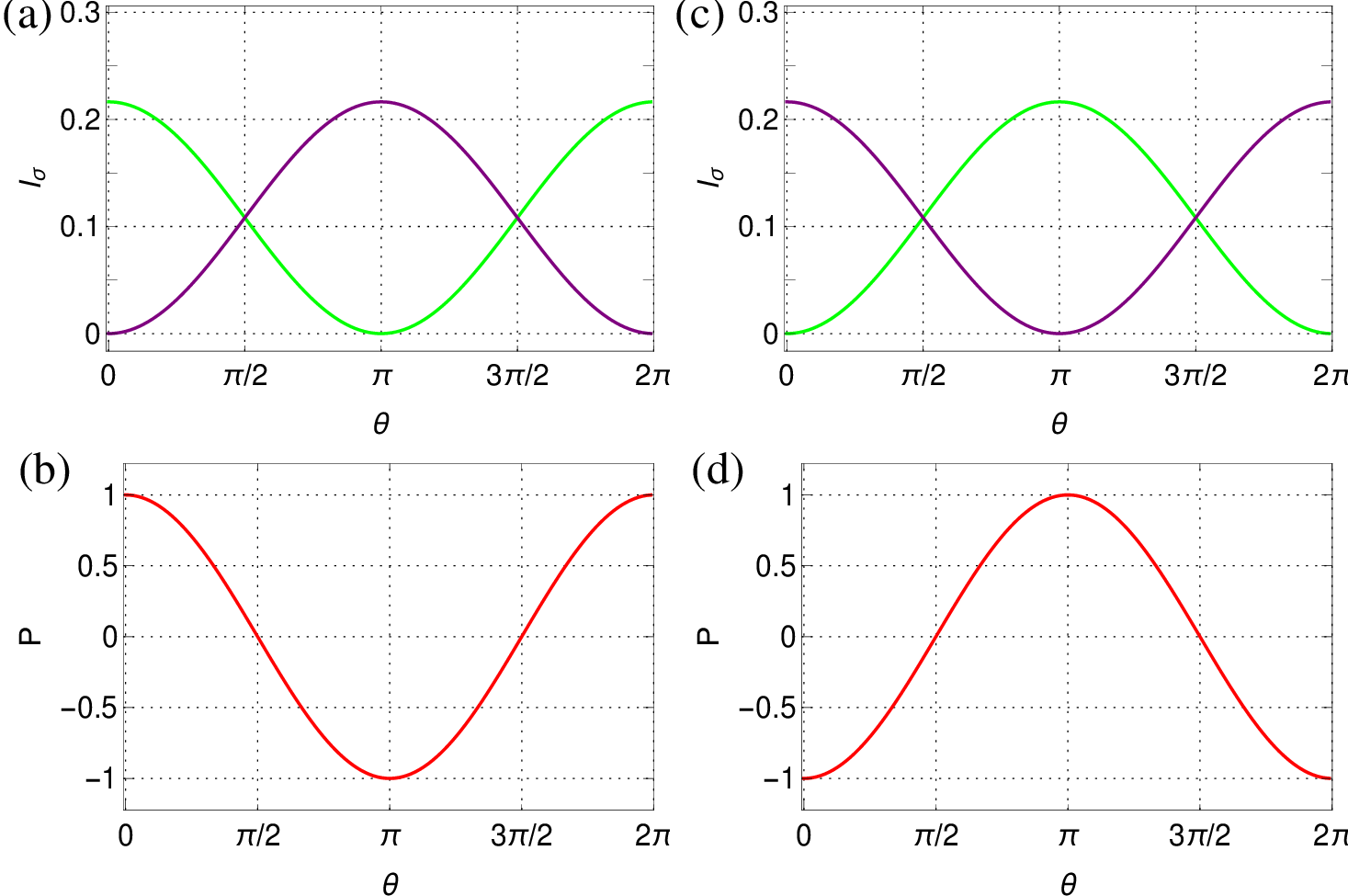}}\par}
\caption{(Color online). Spin-dependent currents and spin polarization as a function of the polar angle of the magnetic moment. 
The green and purple curves in the first row represent up and down spin currents, respectively, while the red curve in the second row 
represents spin polarization. The Fermi energy is set at $E_F = -1$ for (a) and (b), and at $E_F = 1$ for (c) and (d). Other parameters 
are the same as in Fig.~\ref{theta}.}
\label{rotate}
\end{figure}
Thus, polarization reaches its maximum when the polar angle is $\theta = n\pi$. Notably, polarization can be switched from 1 to -1 by adjusting the polar angle from $2n\pi$ to $(2n + 1)\pi$.
\begin{figure*}[ht]
{\centering\resizebox*{14cm}{9cm}{\includegraphics{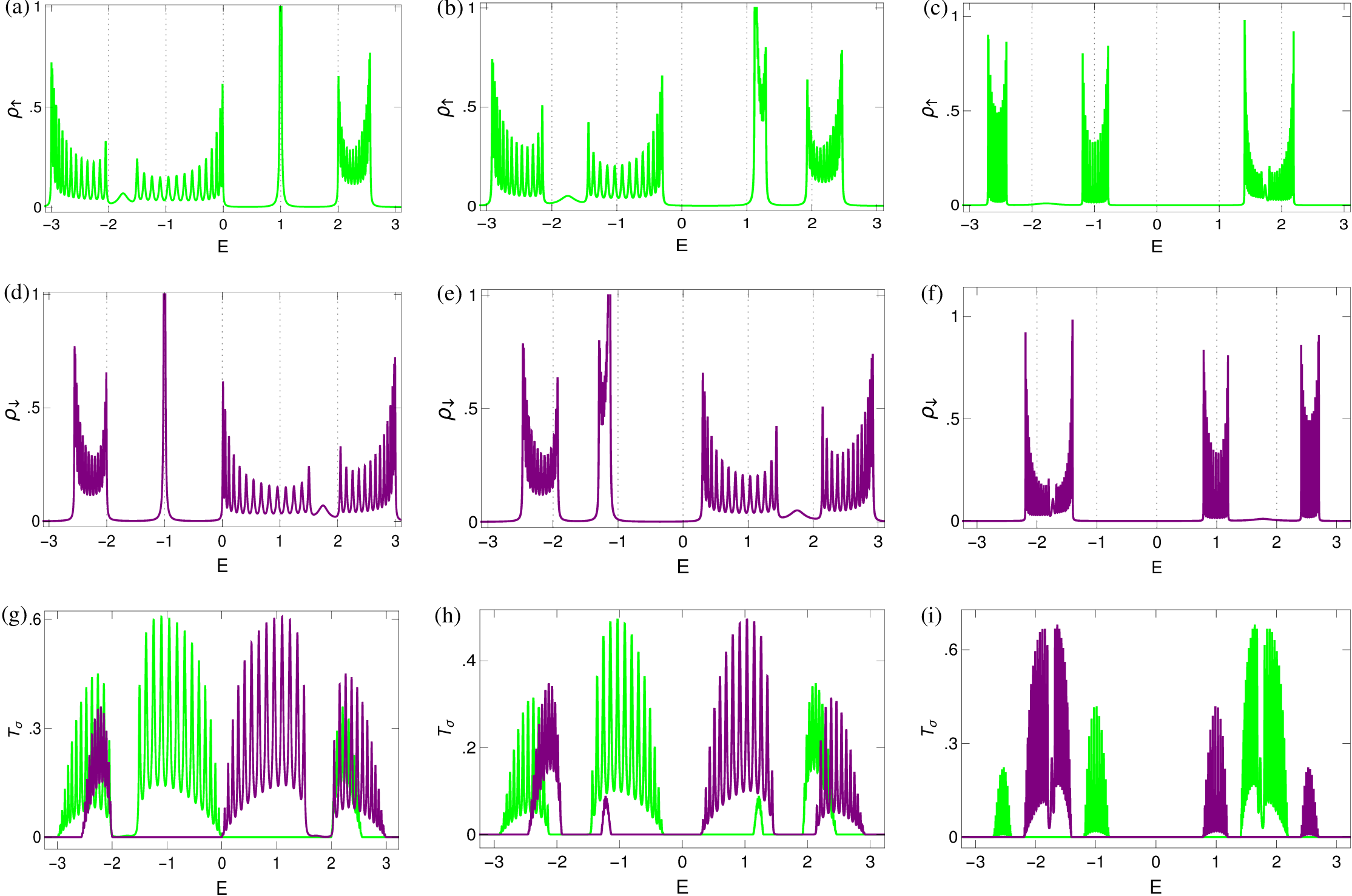}}\par}
\caption{(Color online). Density of states (DOS) and transmission probabilities at different values of magnetic flux $\phi$. Here, the 
spin-dependent scattering factor $h = 1$, and the number of diamond plaquettes is $N = 15$. The green and purple curves correspond to 
up-spin and down-spin electrons, respectively. The first column represents $\phi = 0$, the second column $\phi = 0.2\phi_0$, and the 
third column $\phi = 0.4\phi_0$. Panels (a-c) show the DOS for up-spin electrons, (d-f) for down-spin electrons, and (g-i) the 
transmission probabilities for both spins.}
\label{big}
\end{figure*}

So far, we have discussed spin separation in the antiferromagnetic system in the absence of magnetic flux. The channel mismatch between two different spin-dependent energy channels, and the energy gap between the conducting and localized levels play crucial roles in achieving large polarization. The localized level is $N$-fold degenerate for both spin orientations, but this degeneracy is lifted in the presence of a magnetic flux ($\phi$). We previously examined the effect of magnetic flux in the nonmagnetic case; now, we focus on its impact on the antiferromagnetic system.

\begin{figure}[ht]
{\centering\resizebox*{6.5cm}{7cm}{\includegraphics{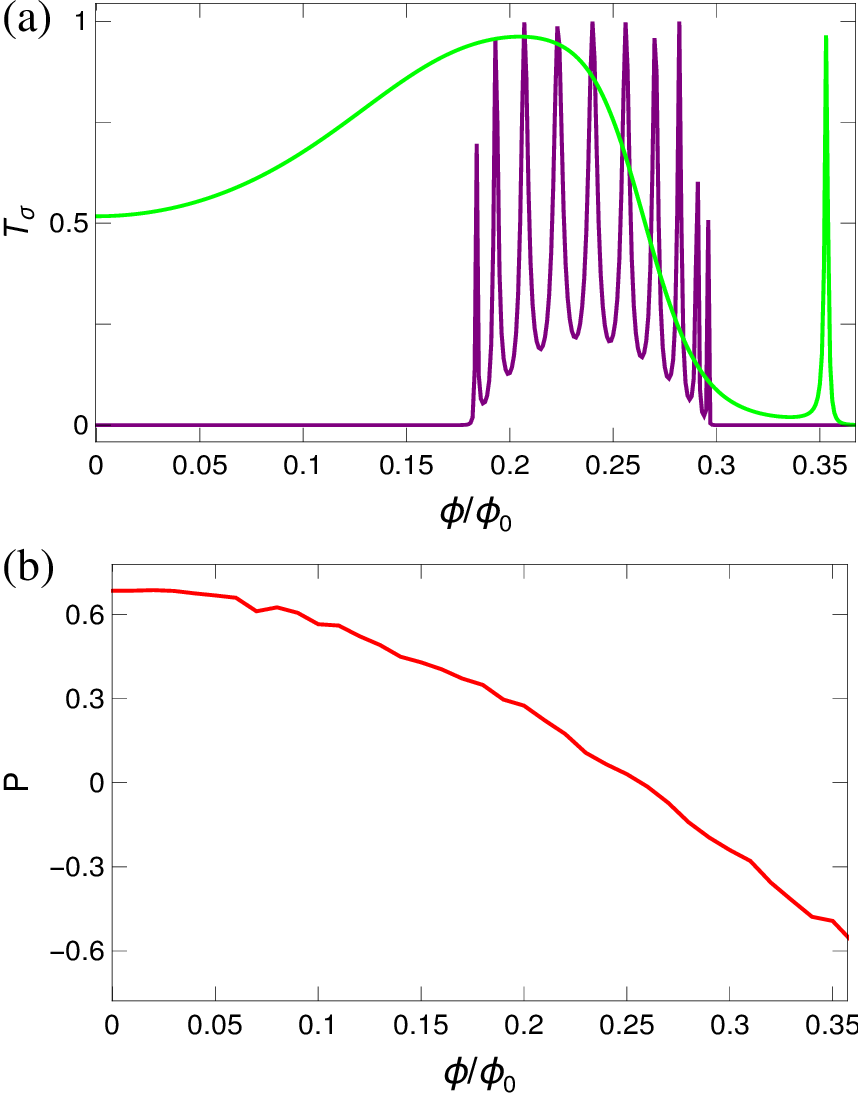}}\par}
\caption{(Color online). A clearer visualization of the specific role of magnetic flux $\phi$. (a) $T_\sigma$ as a function 
of flux, where the green curve represents up-spin and the purple curve denotes down-spin. (b) Variation of spin polarization with 
magnetic flux at $\mathcal{T} = 50\,$K and voltage $V = 2\,$V. The magnetic moment arrangement remains unchanged: 
$\theta = \pi$ for A-type sites and $\theta = 0$ for B-type sites. Here we set $E_F = -1.25$ and $N=15$.}
\label{flux}
\end{figure}
In the absence of $\phi$, a localized energy level appears at $E = 1$ for up-spin electrons, along with three conducting sub-bands at 
different energy levels, separated by a significant energy gap (first row of Fig.~\ref{big}).
\begin{figure}[ht]
{\centering\resizebox*{6.5cm}{7cm}{\includegraphics{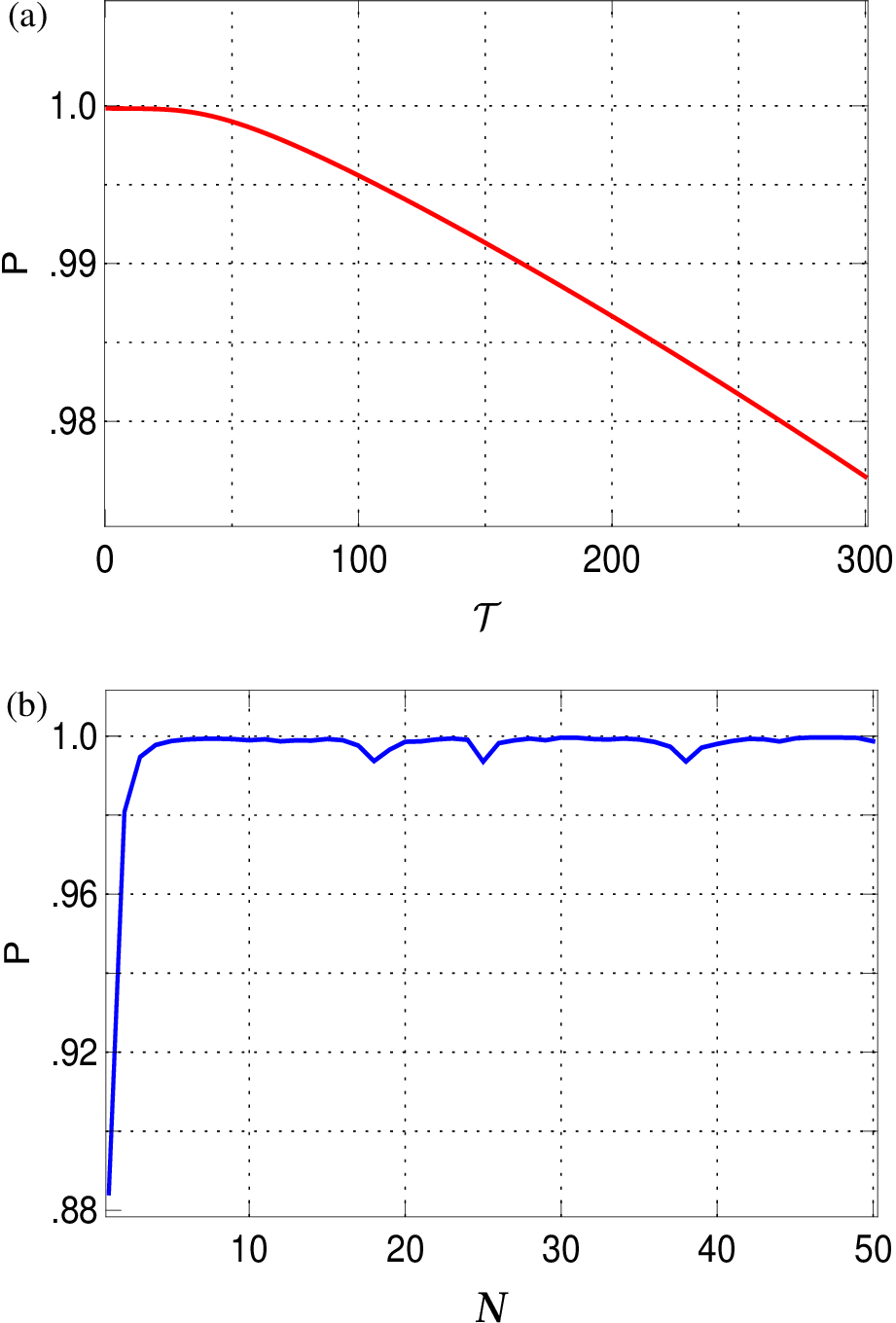}}\par}
\caption{(Color online). Effects of temperature and system size on spin polarization. (a) Polarization versus temperature for $N = 10$
plaquettes. (b) Polarization versus system size at $\mathcal{T} = 50\,$K. In both the cases, we set $E_F = -1$ and $V = 2\,$V.}
\label{system}
\end{figure}
When $\phi$ is introduced, the localized level disappears, and conducting bands emerge in this regime. The energy gap is also affected by 
the presence of magnetic flux. As $\phi$ increases, the newly formed conduction band widens and eventually merges with the nearest 
conduction band. Similar behavior is observed for down-spin electrons, where delocalization occurs at $E = -1$ and overlaps with the 
up-spin energy band (second row of Fig.~\ref{big}). Reflecting the up and down spin DOS spectra, transmission probabilities are obtained for the resonant energy channels
(last row of Fig.~\ref{big}).
At $\phi = 0.5\phi_0$, transmission in both spin channels is suppressed due to complete localization (not shown here in Fig.~\ref{big}), 
similar to the nonmagnetic case. Beyond this, periodicity restores the system at $\phi = 1$.

Figure~\ref{flux} plots the transmission probabilities and spin polarization as a function of $\phi$. For the determining the spin polarization, the Fermi energy set at $E_F = -1.25$. In the absence of $\phi$, up-spin electrons dominate. As $\phi$ increases to approximately $0.18\phi_0$, down-spin transmission begins, eventually surpassing up-spin transmission. This transition is also reflected in the polarization curve, where an initially positive polarization reverses with increasing flux. Thus, the polarity of spin polarization can be tuned simply by adjusting the magnetic flux at a fixed Fermi energy.

In Fig.~\ref{system}, we separately discuss the effects of temperature and system size on spin polarization. We vary the temperature over a wide range while keeping the system size fixed, and then vary the system size while holding the temperature constant. The corresponding results are shown in Figs.~\ref{system}(a) and (b) respectively. Notably, the spin polarization remains stable even at room temperature ($300\,$K), which can be attributed to the significant mismatch between the two spin-dependent energy channels, far exceeding the thermal energy ($k_B\mathcal{T}$). As illustrated in Fig.~\ref{system}(a), the polarization changes by only about $2\%$ over the temperature range $0-300,$K, indicating robust performance at room temperature. On the other hand, Fig.~\ref{system}(b) shows that the polarization remains nearly constant for $N > 5$, suggesting that efficient spin filtration can be maintained across a broad range of system sizes.

\subsection{Semiconductor Properties}

In this sub-section, we discuss how different spin channels can exhibit distinct semiconductor properties by tuning the spin-dependent 
scattering $h$. According to renormalization theory, in the absence of a magnetic flux, a degenerate localized level (LL) appears at $E = \epsilon_A$. For our chosen AFM diamond network (Fig.~\ref{diamond}), the on-site energies for up and down spins are given by $\epsilon + h$ and $\epsilon - h$, respectively.  

When $h = 0$, the LL for both spin channels is located at $E = 0$ (assuming $\epsilon = 0$), meaning there is no band separation between up-spin and down-spin electrons. The energy gap between the conduction band and the LL depends on the on-site potential energy difference between A-type and B-type atoms. In our configuration, the energy difference for up-spin electrons is $\epsilon_A - \epsilon_B = +2h$, while for down-spin electrons, it is $\epsilon_A - \epsilon_B = -2h$. Consequently, at $h = 0$, there is no energy gap, allowing electrons to transmit freely even where LL is located.

In the previous section, we discussed the case of $h = 1$, where a high degree of spin polarization can be achieved. The LL for each spin channel is positioned to open a large energy gap (approximately 1 eV) between the conduction band and the LL. Specifically, the LL for up-spin electrons is located at $E = h$, while for down-spin electrons, it is at $E = -h$.

\begin{figure}[ht]
{\centering\resizebox*{9cm}{9cm}{\includegraphics{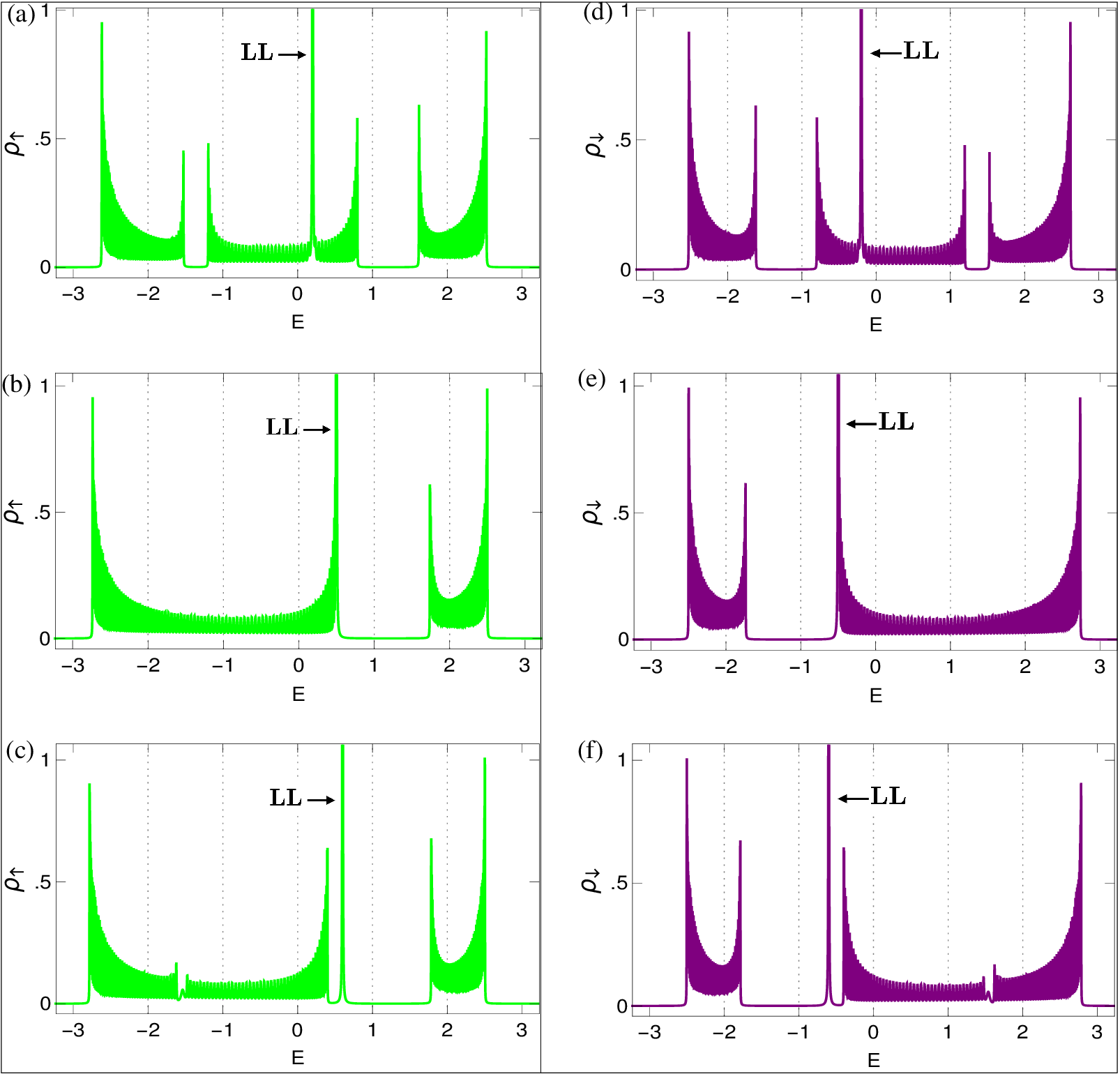}}\par}
\caption{(Color online). Tuning of the localized level by varying the strength of spin-dependent scattering factor $h$. The ADOS for 
up-spin (green) and down-spin (purple) electrons is shown for three values of $h$: (a, d) $h = 0.2$, (b, e) $h = 0.5$, and (c, f) $h = 0.6$. 
The diamond chain consists of $N = 50$ plaquettes. The magnetic flux is set at zero.}
\label{local}
\end{figure}

An interesting scenario arises when the position of the localized level and the energy gap can be selectively tuned. For $h < 0.5$, the LL
resides within one of the conduction bands for both spin channels, as shown in Figs.~\ref{local}(a) and (d). In this case, the LL becomes
insignificant as electrons can transmit through the neighboring conduction band by gaining thermal energy ($k_B \mathcal{T}$, approximately
$0.025\,$eV at room temperature).  

For $h = 0.5$, the LL (see Figs.~\ref{local}(b) and (e)) is located at the boundary of the conduction band. As $h$ increases further, 
the LL moves out of the conduction band, creating an energy gap. At $h = 0.6$, as shown in Fig.~\ref{local}(c), the energy gap between the conduction band and the LL is asymmetric, with a smaller gap on the left side compared to the right. If electrons in the left-side band are excited, 
they transition to the LL. However, once they reach the LL, they become localized. This results in the formation of holes in the left-side 
band, which are up-spin polarized.  

Similarly, in Fig.~\ref{local}(f), the LL for down-spin electrons appears at $E = -0.6$, opening a larger gap on the left side. A small
perturbation can compensate for the energy gap on the right side, allowing electrons in the LL to transition to the conduction band with 
minimal energy input. This results in an $n$-type semiconductor with down-spin-polarized electrons. Thus, by tuning the spin-dependent 
scattering strength and appropriately positioning the Fermi energy, both $p$-type and $n$-type semiconductors can be realized in different 
spin channels.

If an $n$-type semiconductor with up-spin-polarized electrons is desired, a slight modification in the arrangement of magnetic atoms in the diamond plaquette is required. By reversing the magnetic moment direction in A-type and B-type atoms, setting the A-type atom’s moment along the $+z$-axis and the B-type atom’s moment along the $-z$-axis, the spin bands are inverted in energy. In this configuration, the LL for up-spin electrons shifts to $E = -h$, while for down-spin electrons, it moves to $E = h$. Consequently, selecting $h = 0.6$ would produce down-spin-polarized holes and an excess of up-spin-polarized electrons. Therefore, both types of semiconductors can be engineered with either up-spin or down-spin polarization by appropriately configuring the system.

\section{Closing Remarks}

In this work, we introduce an antiferromagnetic model that fulfills two key functions: (i) generating and regulating spin polarization 
and (ii) exhibiting semiconducting behavior, where the spin state of the majority carriers can be controlled. The formation of localized 
levels plays a crucial role in this mechanism. Our model is free from disorder and external electric fields. By simulating the quantum 
system within a tight-binding framework, we characterize its transport properties using Green’s function formalism. All the required 
energy dispersion relations are derived analytically.

A distinctive feature of the energy spectrum is the trio combination of conduction, gap, and localization bands. In low-dimensional systems, 
localized states typically arise from either random or correlated disorder. In the former, Anderson localization~\cite{anderson} occurs 
for any finite disorder, whereas in correlated disorder systems, such as the Aubry-André-Harper model~\cite{aubry,sokoloff,sil1}, localization 
emerges beyond a critical disorder threshold. In contrast, our model exhibits a sharply localized, highly degenerate energy level alongside 
conducting energy channels solely due to the network’s geometry. This unique property allows complete blocking of one spin channel while 
permitting transmission through the other, distinguishing our model within the field.

To ensure a comprehensive understanding, we provide detailed theoretical explanations, making this work self-contained. Additionally, 
various fabrication techniques could facilitate the realization of this diamond plaquette array, including lithographic 
methods~\cite{weekes}, molecular beam epitaxy~\cite{fuhrmann}, and droplet epitaxy~\cite{xia}. The magnetic moment arrangement can be 
controlled via doping with magnetic nanoparticles~\cite{guardia}.

We believe our findings contribute to the advancement of efficient spin-filter devices in small-scale systems and provide a valuable 
example for the expanding field of antiferromagnetic spintronics. 

\appendix
\section{Energy dispersion relations}
\label{app}

We take the nonmagnetic case and numerically solve the E-k relation for three cases. Here the parameters are specified as $\epsilon_A = \epsilon_B = 0$ and $t =t^{\prime} =1$.

\vskip 0.3cm
\noindent
$\blacksquare$ {\bf Solution of dispersion relation for general $\phi$}

\vskip 0.2cm
The four dispersive energy levels correspond to Eq.~\ref{ek1} are given by 
\begin{subequations}
\begin{align}
E_1 = -E_2 = \frac{1}{2} \left(\alpha+\beta\right) \label{ap1} \\
E_3 = -E_4 = \frac{1}{2} \left(\alpha-\beta\right) \label{ap2}
\end{align}
\end{subequations}
where $$\alpha = \left[\frac{1}{q} \left(20.57-10.07 \cos{4 \eta}\right)+ 0.26 q + 3.33\right]^{1/2}$$ 
and $$\beta = \left[10 - \alpha^2 - \frac{8}{\alpha} \cos{2 \eta} \cos{k} \right]^{1/2}.$$
The factor $q$ is expressed as
\begin{eqnarray*}
q = \left[470 -720 \cos{4 \eta} + 432 \cos^2{2 \eta} \cos^2 {k}  \nonumber\right.\\
\left.+\left\{55296 \left(\cos{4 \eta} -2.04\right)^3 + 518400 \left(0.65 \nonumber\right.\right.\right.\\
\left.\left.\left.-\cos{4 \eta} + 0.6 \cos^2{2\eta} \cos^2{k}\right)^2\right\}^{1/2}\right]^{1/3}.
\end{eqnarray*}

\vskip 0.3cm
\noindent
$\blacksquare$ {\bf Solution of dispersion relation for $\phi = 0$}

\vskip 0.2cm
Here we get one localized level and three dispersive levels as 

\begin{subequations}
\begin{align}
E_1 = 0 \label{ap3} \\
E_2 = 3.46 q_0^{-3} + 0.48q_0 \label{ap4} \\
E_3 = -\frac{1}{q_0}\left(1.73 - 7 j\right) -q_0\left(0.24 + 0.42 j \right) \label{ap5} \\
E_4 = -\frac{1}{q_0}\left(1.73 + 7 j\right) -q_0\left(0.24 - 0.42 j \right). \label{ap6}
\end{align}
\end{subequations}

\vskip 0.3cm
\noindent
$\blacksquare$ {\bf Solution of dispersion relation for $\phi = 0.5 \phi_0$}

\vskip 0.2cm
Here we get four localized energy levels as 
\begin{subequations}
\begin{align}
E_1 = -E_2 = 1 \label{ap7}\\
E_3 = -E_4 = 2. \label{ap8}
\end{align}
\end{subequations}

\section*{DATA AVAILABILITY STATEMENT}

The data that support the findings of this study are available upon reasonable request from the authors.

\section*{DECLARATION}

{\bf Conflict of interest} The authors declare no conflict of interest.

\end{document}